# Learning-focuses in physics simulation learning situations

Åke Ingerman, Cedric Linder[*], and Delia Marshall

*Department of Physics, University of the Western Cape, South Africa, and [*]Uppsala University, Sweden.*

aingerman@home.se; cedric.linder@fysik.uu.se; dmarshall@uwc.ac.za

This is a report on a qualitative study of students' learning where a physics computer simulation session is used to supplement lectures on the topic. Drawing on phenomenography as the analytical framework, the students' learning-focuses were analysed. The result is a description of four distinctly different learning-focuses that emerged when the students involved in the study interacted with the computer simulations. These learning-focuses were then analysed in terms of the level of interaction, the nature of physics knowledge and views of learning experienced by the students. These results were then used to identify advantages and disadvantages of learning through interaction with simulations.

**Introduction and background**

There is a growing recognition that the passiveness inherent in much traditional undergraduate physics teaching often contributes to poor learning outcomes (for example, see Redish & Steinberg, 1999). Hence undergraduate physics teaching is increasingly drawing on active supplements to traditional teaching, such as computer simulations (Christian and Bellani, 2001; Bates, 2002). However, there has been little associated research, which problematizes this kind of supplemental teaching from the learners' perspective. This paper reports on part of a project that is aimed at addressing this shortfall in terms of characterizing physics learning from simulations. We characterize the learners' experiences of the process of learning through the means of a simulation of a physics phenomenon, as 'learning-focuses'. With 'learning-focus' we mean to describe the kind of (or class of) features which are apparent for the students in the learning situation and thus are dominating the students' experience of it. To put it more theoretically, with 'learning-focus' we mean to capture what features constitute the students' experience, describing the distinctive quality these features have (and also implying what is taken for granted and what is in focal awareness for the learners).

These learning-focuses are at a general level. What is not captured by such a description is a characterisation of learning at a micro-level, i.e. the specific learning around the particular physics phenomena represented by the simulation. In other words, a description of what important aspects of the phenomena themselves are brought into students' focal awareness during the simulation session, and how this occurs. However, the investigation at the general level opens up the path to such investigations at the micro-level (an aspect of our research project not reported here), while still being interesting in itself.

**Analytic framework and data collection**

Our intended research question shifted during the study. We started out trying to describe learning on a micro-level. This was done by drawing extensively on the phenomenographic research tradition and its anatomy of awareness theory as sketched out by Marton & Booth (1997), and its applications as exemplified by Marton & Morris (2002). In particular, constructs such as focal awareness, intended and enacted object of learning, and variation of critical features, were important. We collected data in different situations, and in different ways. While trying to organise and analyse our data, some structure emerged about the apparent features the students saw in the different learning situations, showing similarities on the general level. We followed this strand of analysis, seeing that these apparent features (now developed into learning-focuses) might offer insights into learning situations aided by a simulation, both on the micro-level and a general level.

In this regard the analysis can be characterised as being an extensive qualitative exploration of *variation in meaning* that learners see in particular simulation situations and in the physics phenomenon the simulations are representing and presenting. The outcomes described emerged from analysis of video-audio observations and interview descriptions centred around students' interactions with physics simulations in areas of *electric potential, field and force* (introductory level), the *Bohr model of the atom* (introductory level), and the *quantum mechanics scattering of a wave function at a barrier* (senior level). The participating students - physics undergraduates - were from a South African and a Swedish university. Hence the data collection was situated in different levels of learning complexity (introductory and advanced undergraduate physics), different physics concepts and phenomena, different prior learning experiences with both simulations and computers (from being the first encounter with simulations to having created simulations themselves), and in different cultural settings (South Africa and Sweden).

**Results**

The analysis led to the identification of four qualitatively different learning-focuses used when interacting with one of the given physics simulations. This variation in learning-focus we characterised as follows:

*Learning-focus A - Focusing on simulations as given assignments*

Framing a simulation session with learning-focus A implies being mainly concerned with the constraints and features of the situation as an assignment that needs to be completed. External demands and expectations dominate the experience, while the simulation and the phenomenon it represents are not really discerned at all, with respect to their meaning.

In this transcript, two students are working with a Bohr's model simulation[1], going through the 'tutorial' questions that come with the simulation, and are presently dealing with Tutorial Question 2 (Q2). Before starting, they had been instructed to explore the simulation, guided by the questions, which were to be used as a background when discussing their experience of working with the simulation. In the simulation, the students have a visual presentation of an atom with circular orbits with an electron going round in one of the orbits. The simulation also provides them with a diagram of the orbits' corresponding energy levels (E1,...,E6). The students can initiate transitions between these levels and see to which spectral lines the transitions correspond.

> *S1:* [Reads Q2 aloud.] *"How will the wavelength of the emitted photon, as the electron returns to the ground state, compare with the wavelength of the absorbed photon, which originally excited the electron to the 5th orbit?" What?!?* [Seems puzzled.]
> *S2:* [Having read Q2 silently again.] *I don't understand this.*
> *S1: Okaaay.... I understand. We first have to take it to the 5$^{th}$ orbital, and then return to the ground state, comparing the wavelengths.* [They move the electron from the ground state (E1) to the 5$^{th}$ orbit (E5).]
> *S2: So, now we take it back to the ground state.* [They do E5 to E1.] *There's no change!*
> *S1: Is it? Do it again.* [Sounds doubtful.]
> *S2:* [Repeats transitions.] *There's no change.*
> *S1: Okay, there's no change. Now go to ground.*

Here all that matters is that they get to an answer -- there is no discussion of why they observed what they did. They then spent a long time writing down their answer, followed by a discussion about which question to do next.

> *S1: She* [the interviewer] *said we must leave out Q4 and Q6.*
> They proceed to Q3; do that in rather a perfunctory way. The only discussion is about how much time they have left to complete the simulation tasks and about an assignment due later in the day.

---
[1] Taken from *ActivPhysics* by van Heuvelen A & D'Alessandris, 1999, Addison-Wesley.

> *S1: Okay, onto Q3.*
> *S2: We've done Q3. This* [S2 points to chunk of text.] *– this they just tell you. Now, let's go down….*
> They omit the important part of the text in Q3, instructions to check the correctness of their answer to Q3 by checking that they did indeed produce the wavelength on extreme right of spectrum. They then press on to finish.

The focus of the students is on what they think is expected by the interviewers from working with the simulation. They seem to be mainly concerned about getting to an answer and not on achieving a learning outcome.

*Learning-focus B - Focus on the presentation*

In this learning-focus the simulation is distinctly in the foreground -- learning is primarily about the simulation. The physics situation the simulation is meant to illustrate and represent is not really part of the experience, apart from the simulation being seen as a visual representation, which can be taken at face value. The relationship to the learning situation is passive, the simulation session is experienced as a demonstration that can be seen, possibly explained, and understood.

This transcript is from the same session used to illustrate learning-focus A, but occurred when the students were working with Question 1, still being not totally familiar with the simulation:

> *S1:* [Reads Question 1.] *"Does it take more energy for the electron to jump from the ground state to the 2nd orbit or from the ground state to the 3rd orbit?"*
> *S1: Okay, so if you want the electron to move to the 2nd level, it needs energy.*
> *S2: How do you know that?*
> *S1:* [Reads aloud.] *"In order for the electron to occupy a higher energy state, it must receive energy…"*. *Meaning that if you want to move the electron to the 2nd energy level, it requires energy. Do you understand?*
> *S2:* [Looks at the screen.] *What level is it in at the moment?*
> *S1: It's currently at E3.* [Points to E3 on energy level diagram.] *That's E1* [Points.] *… E2* [Points.] *… E3.* [Points.]
> *S2: Meaning this is E1?* [Points to innermost orbit on orbital diagram.] *Then E2, E3?* [Points.] *Okaaay!* [Aha!]

The student S2 after a while realises that a relationship exists between the energy level diagram and the orbit level diagram shown in the simulation. S1 helps by demonstrating and explaining: In contrast to the transcript illustrating learning-focus A, where we interpreted that the two students' learning-focus was the same, here only student S2 uses learning-focus B.

*Learning-focus C - Focus on manipulating the simulation*

Here the simulation is actively used to understand something. Through the means of the simulation, the underlying physics phenomenon (or 'reality'), which is taken to be one with the simulation, can be understood. However, the simulation is the primary focus. Unlike learning-focus B, here the learner is the active agent choosing what is to happen (and be learnt), and is not just a passive observer.

This transcript is from an interview with two students on their experience of working with a simulation of electric potential, field and force.

> *S1: Because when you actually - it's like an experiment, the simulation itself is like an experiment, because when you actually do it, then it actually clicks in your mind already, and then after you go through the text it's sort of like 'ag, I know that'.*
> *S2: Yes, for instance when you have point charges, because when you made one bigger - I mean, you played already with them - you made one bigger and you could see the force*

> *changing, and you move them apart and you could see the arrows changing, that sort of thing. And then you read it* [the explaining text that goes with the simulation] *afterwards and it was just…an afterthought.*

Here an appreciation may be obtained about how – through the simulation – the students can control the physics phenomenon and thus feel able to claim they understand it.

*Learning-focus D - Focus on exploring the phenomenon*

Using learning-focus D implies an awareness of the representational nature of the simulation, and of the limits of both the simulation and the phenomenon as tools to understand the world. The simulation is used as a conceptual tool, which can be used to explore and understand the phenomenon. On the one hand, learning about the simulation happens in relation to the phenomenon it is meant to represent and on the other hand, learning about the phenomenon happens with the limitations of the simulation in mind.

In this transcript, a student is describing working with a simulation on scattering of an electron wave packet on a potential barrier in quantum mechanics -- working with the simulation is a self-constructed exercise with the clear aim of exploring the phenomenon, giving access to exploring aspects of the phenomenon not easily accessible otherwise.

> *We tried a lot and ... looked at it and we sat for hours, just tried different potentials and, somewhere, that interest,* [we] *must have had an interest in doing that, otherwise we wouldn't have spent an entire Sunday just sitting with different potentials. And our project was finished more or less, but we couldn't let go, so there I think, it looks nice, its quite funny with these little things that are flopping around and as it comes to understanding and relevance to what we were doing in the actual course .... And as I said earlier, when you think about electron configuration you don't really have to, you never have to interpret the wave function as a particle somewhere, it is more that you have a static solution that is, you don't really have to think about what it is you have. A model that someone has thought up, and you have experiments verifying them. This model works pretty good, and since we are not able to see it, we will never be able to know exactly what it is and then it is fine and you calculate with it...*

The student describes a process of trying things out, pondering about the nature of the phenomenon of the wave function as representation of the electron, how it is shown in the simulation, and how they could be used to understand more about the behaviour of electrons. The simulation is taken to be a good representation of a model, which someone thought up, being related to 'reality' through experiments, while still recognising the non-tangible nature of this type of physics phenomena.

**Discussion**

In analysing the learning-focuses, trying to bring out more of their essence, we found it useful to analytically differentiate between three different levels of learning interaction: the level of the situation, the level of the simulation and the level of the phenomenon.

    The *level of the situation*. This analytic level is about the *setting* of the situation in terms of what is seen to be relevant (by the learner) for the associated learning. For example, seeing a situation as being similar to other situations already experienced, general perceptions of what it means to learn physics and work with a computer, and the experienced demands and expectations – how working with the simulation is understood as a learning situation.

    The *level of the simulation*. This analytic level reflects the kinds of understandings formed about how a simulation 'works' and thus what the learners see as being prominent features of the simulation, and the logical and conceptual links between the different aspects of a given simulation.

    The *level of the phenomenon.* This analytic level is about what is experienced by learners when they create conceptual links between the simulation and the represented phenomenon.

> For example, learners need to have some notion of a phenomenon in order to relate to what the simulation is representing.

Of course in the experience of a given simulation all of these levels are interrelated, but at a particular time, learners working with the simulation maybe focally aware of all three levels in quite different ways. In this regard, an analysis of learners' *learning focuses* in terms of these three levels was made. Here it is important to point out that if focus is found to be directed primarily to one level it cannot be concluded that there is no awareness or response to the other levels. But it does provide a good insight into where the essential learning interaction is taking place.

We found that the learning-focuses could be characterized as having a main level of interaction. Further, we could see two other important aspects of the learning-focuses: the view of learning and the view of the nature of physics implied. The analysis in terms of these three aspects is summarized in Table 1.

Table 1: Overview of the essence of different aspects of the learning-focuses.

| Learning focus: (Focus on) | *Main level of interaction* | *Nature of physics knowledge* | *View of learning physics* |
| --- | --- | --- | --- |
| **A - the situation** | situation | physics not really present | fulfilling demands, tasks |
| **B - the presentation** | simulation | physics phenomenon to be seen and explained | learning by seeing and someone explaining |
| **C - manipulating** | simulation | physics phenomenon to be controlled | learning by doing yourself and understanding |
| **D - exploring** | phenomenon | ways of seeing, predicting and interpreting phenomenon | explorative creation of a body of knowledge |

Obviously all the learning-focuses are not as productive in terms of learning. Learning focus A is non-productive, while we see different degrees of value in the others. Learning focus B and C have in common that their main level of interaction is the simulation, and the simulation is taken as a window into 'reality'. Learning focus C presupposes an understanding of how the simulation itself works, thus implying that learning focus B at some point has been used. However, employing learning focus B does not imply that learning focus C will be taken. In most of the simulation sessions we studied, we characterised the learning focus as B or C. Learning focus C gives the possibility of developing some new understanding about the physics phenomenon. However, what can be learnt is restricted in that the simulation is taken at face value and, for example, what is not included in the simulation is not considered. If only learning focus B is used, the learning outcome is even more restricted and comparable to what a video or demonstration can 'tell' the students.

Learning focus D in some ways sidesteps the other learning-focuses in that here the difference in the view of learning and nature of physics knowledge implies different ways of engaging with the simulation. The view implied in learning-focuses B and C - that the simulation should and could be taken at face value - is not compatible with the recognition in learning focus D that both the simulation and the underlying phenomenon are representational in nature, having limitations in their validity and applicability. The learners' epistemological beliefs and views of learning can make creating and retaining learning focus D very difficult, and severely restrict what is possible to be learnt. It is not surprising that we did not observe any introductory students taking such a focus, since introductory students are unlikely to have developed such a sophisticated view of physics. Even though the change in the view of physics required to adopt learning focus D instead of B and C must be part of a broader learning about physics, the simulation situation offers students some extra possibilities of not only learning about the phenomenon itself but also of developing their views of physics or views of learning physics, in particular if supported by a mindful and attentive teacher.

What then are the implications of the learning-focuses for a teacher employing simulations as a support in creating good learning contexts for students? Clearly from our data, the level on which we can expect students to engage is with learning-focus B and C. With these learning-focuses, interventions are important and can be catalysts for the learners' learning. For learning through simulations, things taken for granted need to be brought into focal awareness. This might happen by students spontaneously bringing two aspects presented by the simulation together, being shown by the simulation and observed by the students, or questioned by an intervener, for example a fellow student or a teacher. From the analysis of all our data, we conclude that a *mindful* intervention by a teacher, juxtaposed with the simulation possibilities, is a powerful tool for learning. In itself, the learning possibilities in a simulation session can easily be glossed over by the learner. Of course, simulations working as a powerful tool for learning presupposes that the simulation has relevant features, that the learning tasks are manageable and seem relevant to the students, and that interventions are mindful - in short, that the simulation learning context shows attributes of 'good teaching', as characterised by research (see for example, Ramsden, 1992). However, we recognise that learners working with learning-focus D could probably meaningfully be working on their own, setting their own learning agenda, and taking the responsibility of making mindful use of the simulation's learning possibilities.

We can also conclude that the passiveness inherent in much traditional undergraduate physics teaching is not necessarily eliminated by the use of interactive simulations, but that simulation learning situations can also support poor learning outcomes. However, simulations can be a powerful pedagogical tool for bringing central features of a phenomenon into focal awareness for students, creating wider possibilities for learning. Metaphorically, the space of potential learning is spanned by the students' learning-focus, by the possibilities of showing variation in important aspects of the phenomenon in the simulation, by the interveners' mindfulness and by the appropriateness of the learning tasks.

**Acknowledgements**

This project has been financially supported by the South African National Research Foundation (NRF), the Swedish International Development Cooperation Agency (SIDA), the Swedish Research Council (VR), and the Swedish Foundation for International Cooperation in Research and Higher Education (STINT).